\input{aipcheck}

\documentclass[sort,
    ,final            
  ]
  {aipproc}

\usepackage{amssymb}

\layoutstyle{8x11double}

\begin{document}

\title{Higher Dimensional Operators in the MSSM\footnote{This
is based on the talks of the authors (I. A. and D.M.G.), 
presented at the Planck 2008 conference (19-23 May, Barcelona),
and SUSY 2008 (16-21 June, Seoul), and it will appear in the
proceedings of the SUSY~2008 conference. }}


\author{I. Antoniadis}{
  address={Department of Physics, CERN - Theory Division, 1211 Geneva 23,
 Switzerland.},altaddress={CPhT,  \'Ecole
 Polytechnique, CNRS, 91128   Palaiseau Cedex, France.}
}

 \author{E. Dudas}{
   address={CPhT,  \'Ecole Polytechnique, CNRS,  91128
   Palaiseau Cedex, France.},altaddress={
   LPT, B\^at 210, Universit\'e de Paris-Sud, CNRS,
  91405 Orsay Cedex, France.}
 }

\author{D. M. Ghilencea}{
  address={Rudolf Peierls Centre for Theoretical Physics, 
University of Oxford, 1 Keble Rd, Oxford OX1 3NP, U.K.}
}

 \author{P. Tziveloglou}{
   address={Department of Physics, Cornell University, Ithaca, NY
  14853 USA.}
,altaddress={CPhT,  \'Ecole
 Polytechnique, CNRS, 91128   Palaiseau Cedex, France.}
 }

\begin{abstract}
The origin and the implications of higher dimensional
effective operators in 4-dimensional theories
are discussed in non-supersymmetric  and supersymmetric cases.
Particular attention is paid to the role of general, derivative-dependent
field redefinitions which one can employ to obtain a simpler form 
of the effective Lagrangian.
An application is provided for the 
Minimal Supersymmetric Standard Model extended with 
dimension-five R-parity conserving
 operators, to identify the minimal irreducible set of 
such operators after supersymmetry breaking.  Among the physical
consequences of this set of operators
are the presence of corrections  to the MSSM Higgs sector and 
the generation of ``wrong''-Higgs Yukawa couplings and 
fermion-fermion-scalar-scalar interactions. These couplings
have implications for  supersymmetry searches at the LHC.
\end{abstract}

\maketitle

\subsection{Introduction}

The Standard Model (SM) and its  minimal supersymmetric 
version (MSSM)  are the best models we currently have for 
describing the low-energy physics. Despite their success, 
there are many reasons to think 
that they are only a  low-energy manifestation
of a more fundamental theory (supergravity, strings, extra dimensions, etc),
 valid at   higher scales. However, the exact details 
of such a fundamental theory  are in many cases unknown 
(moduli problem, vacua degeneracy, etc) and making 
definite predictions for new physics is 
 difficult. One possibility to investigate {\it new physics} beyond
the  SM/MSSM is to  use instead an effective 
field theory approach. This approach
is a fully consistent and useful  framework  for such study.

In effective field  theories,  operators of dimension larger than four
are present, suppressed by the (high) scale of new physics $M_*\gg  m_Z$.
The origin and the effects of these operators
are discussed in this talk, for the case of 4 dimensional
non-supersymmetric and supersymmetric theories.
We shall distinguish two classes of higher dimensional operators:
\emph{class A} of operators
 involving at most two 
 space-time derivatives acting on physical fields (one derivative 
in the case of fermions); 
\emph{class B} of operators which contain more than two
such derivatives (one in the case of fermions). In general these classes of
operators are not entirely independent of each other.

Regarding their origin,  
higher dimensional operators are generated by integrating out
new physics at $M_*\gg m_Z$ ($M_*\sim$ TeV or higher).
In compactification of higher dimensional theories such 
operators are usually generated, suppressed by the volume of
compactification. Much more 
commonly, higher dimensional operators 
are generated in 4D \emph{renormalisable} theories, after integrating
out massive states. Therefore, although some interactions
may look non-renormalisable in the effective  formulation, 
they may actually be a low energy manifestation of a renormalisable theory
valid up to a much higher scale. The familiar Fermi interaction
 is such  an example.

The power of the effective approach resides in arranging these
operators in series of powers of $1/M_*$, to which 
additional organising criteria, such as symmetry arguments 
inspired by low-energy phenomenology, are also considered. 
The effective Lagrangian has then the form
\begin{equation}
{\cal L}={\mathcal L}_{0} +\sum_{i, n} \frac{c_n^i}{M_*^n}
\,{\mathcal O}_n^i
\end{equation}
where ${\cal L}_0$ is the SM or the MSSM Lagrangian;
${\cal O}^i_n$ is an operator of dimension $d=n+4$ with the index 
$i$ running over the set of operators of a given dimension, and 
$c_n^i\sim {\cal O}(1)$.
This description is appropriate for scales $E$ which satisfy
$E\ll M_*$. Constraints from phenomenology can then be used
to set bounds on the scale of new physics $M_*$. 
The effects of ${\cal O}_n^i$ on observables
can be comparable to one-loop effects in
the SM/MSSM, as we shall see in an example, and this
shows the importance of their study.

\subsection{The non-supersymmetric case}

Let us see some examples of the origin of these operators.
 Consider first the case of
a tree level exchange 
of a massive $Z'$ gauge boson beyond the SM or
MSSM:
\begin{equation}
{\cal L}\supset \big\vert
(\partial_\mu-i\,Z_\mu')\,H\big\vert^2-\frac{M_*^2}{2}
\,Z_\mu^\prime\,Z^{\prime  \mu}
\end{equation}
 After  integrating out $Z'$,  a higher dimensional
 operator of class~A is generated for instance for the Higgs field $H$, 
 which we denote $\Delta{\cal L}$:
\begin{equation}
\Delta {\cal L}=
 \frac{1}{M_*^2}\,(H^\dagger \partial_\mu\,H)^2\qquad
\end{equation}
Similarly, for fermions charged under $Z'$, one finds
\begin{eqnarray}
{\cal L}&\supset&
i\,\overline\psi\, \gamma^\mu\,D_\mu\,\psi
-\frac{M_*^2}{2}\,\,Z_\mu'\,Z^{' \mu}
\nonumber\\
\Rightarrow\quad\Delta {\cal L}&=&
\frac{1}{2\,M_*^2}\,(\overline \psi \gamma_\mu\psi)^2
\end{eqnarray}
There are also operators of class B which can be generated, by the kinetic
mixing of light 
with
massive states, upon integrating out
the latter. For example, from
\begin{eqnarray}
{\cal L}&=&\frac{1}{2}\,(\partial_\mu\phi)^2
+\frac{1}{2}\,(\partial_\mu\chi)^2
+c\,\partial^\mu\phi\,\,\partial_\mu\chi
\nonumber\\
&-&\frac{\lambda}{4}\phi^4
-\frac{1}{2} M_*^2 \chi^2-\frac{1}{2}\lambda^\prime\phi^2\chi^2
\end{eqnarray}
one finds upon integrating out the massive field $\chi$:
\begin{eqnarray}
{\cal L}&=&\!\!\frac{1}{2}\,(\partial_\mu\phi)^2
-\frac{\lambda}{4}\phi^4
+\frac{c^2}{2}\,\Box\phi\,\frac{1}{M_*^2+\Box+\lambda^\prime \phi^2}
\,\Box\phi\nonumber\\[5pt]
&=&
\frac{1}{2}\,(\partial_\mu\phi)^2
-\frac{\lambda}{4}\phi^4
+\frac{c^2}{2\,M_*^2}\,(\Box\phi)^2+\cdots\label{truncation}
\end{eqnarray}
This contains higher dimensional operators of class B 
(more than two derivatives).
If one truncated the series of derivatives to the last term in 
the second line in (\ref{truncation}), 
after a field redefinition one obtains a formulation of 
${\cal L}$ which involves only two space-time derivatives 
but negative metric (ghost) fields
\cite{Hawking}. Obviously, this is an artifact of the truncation:
there are no ghosts present in the theory, 
as long as one retains the whole series of expansion 
in derivatives and provided that the original theory
was ghosts-free. 
We  shall generalise this to the supersymmetric case
(see \cite{Antoniadis:2007xc}).

Higher dimensional operators of class B are also present in 
the low energy effective action of 
string theory.
One can expand the Dirac-Born-Infeld 
 action, to find an infinite 
series of such operators.  Also $\alpha'$ and loop corrections
in string theory generate higher derivative
operators. In effective field  theories of compactification
such operators are generated dynamically 
at one-loop \cite{NH1,NH2,Gh0,Gh1,Gh2} after integrating
out momentum modes. 

\subsection{The supersymmetric case}

Consider a general 2-derivative supersymmetric Lagrangian, with
the following functions of the chiral superfields $\Phi_i$:
the K\"ahler potential $K$, the superpotential $W$ and the 
gauge kinetic function $f$:
\begin{eqnarray}
{\cal L}&=&
\int d^4\theta \,K(\Phi^\dagger_i\,e^V, \Phi^i)\nonumber\\
&+&
\int d^2\theta \Big[W(\Phi_i)+f_{ab}(\Phi_i)\,\,{\cal W}^a\,{\cal W}^b
\Big]+h.c.\qquad 
\end{eqnarray}
where ${\cal W}^a$ is the supersymmetric gauge field strength
associated to the vector superfield $V^a$.
The presence of higher dimensional operators is hidden in the
power expansion (in fields) of these functions:
\begin{eqnarray}
K&=&\Phi^\dagger_i\,e^V\,\Phi^i+\bigg[\frac{c^i_{jk}}{M_*}\,\Phi_i^\dagger
e^V \Phi^j\Phi^k+h.c.\bigg]
+\cdots\qquad
\nonumber\\
W&=& \lambda_{ijk}\,\,
\Phi^i\Phi^j\Phi^k+\frac{c_{ijkl}}{M_*}\,\,\Phi^i\Phi^j\Phi^k\Phi^l+\cdots
\nonumber\\
f_{ab}&=& \delta_{ab}+\frac{f_{abi}}{M_*}\,\,\Phi^i+\cdots
\end{eqnarray}
The first term in the rhs would lead to a renormalisable theory.
These functions can contain not only operators of class A,
but also operators of class B. For example one can have:
higher derivative operators in the superpotential (a) and in
the K\"ahler potential (b):
\begin{eqnarray}
\!\!\! (a)\,\,\,\,\frac{\lambda_{ij}}{M_*}
\int d^2\theta\,\Phi_i\,\Box\,\Phi_j \,\sim \frac{\lambda_{ij}}{M_*}
\int d^4\theta \,\Phi_i\,D^2\,\Phi_j\qquad\,\,
\nonumber\\
\!\!\! (b)\,\,\,\frac{k_{ij}}{M_*^2}
\int d^4\theta\,\,\Phi^\dagger_i\,\Box\,\Phi_j,\,\,\,
\frac{k_{ij}}{M_*^2}\,\int d^4\theta\,\,
\Phi^\dagger_i\,\Phi_j\,D^2\,\Phi_k,...
\end{eqnarray}
where $D$ is the chiral supercovariant derivative.
In  $(a)$, terms like $\psi\Box\psi$ and  $F\Box\phi$ are
generated, where $\Phi=\phi+\sqrt2\theta\,\psi+\theta^2\,F$.
In  $(b)$ one finds terms like $\vert \Box\phi\vert^2$,
$\overline\psi\partial_\mu\Box\psi$, $F^\dagger \Box F$.
Therefore for class B operators, auxiliary fields become 
dynamical degrees of freedom.

Let us present the origin of such operators in a simple case of
a 4D supersymmetric {\it renormalisable} theory.
Consider the Lagrangian
\begin{eqnarray}
{\cal L}&=&
\int d^4\theta\,\Big[\Phi^\dagger\Phi+\chi^\dagger\chi\Big]
\nonumber\\
&+&
\int d^2\theta \Big[\,m\,\Phi\chi+\frac{M_*^2}{2}\,\chi^2+
\frac{\lambda}{3}\Phi^3\Big]+h.c.\qquad
\end{eqnarray}
Using the eqs of motion of the massive field $\chi$ and some 
field redefinitions \cite{Antoniadis1}, one obtains
\begin{eqnarray}
\!\!\!{\cal L}&=&\!\!\!\!\!\!
\int d^4\theta\,\Big[
\Big(1+\frac{m^2}{M_*^2}\Big)\Phi^\dagger\,\Phi+
\frac{m^2}{M^4_*}\,\Phi^\dagger\Box\Phi+\cdots\bigg]
\nonumber\\[5pt]
&+&\!\!\!\!\!\!\!\!\int d^2\theta\,
\bigg[\frac{-m^2}{2\,M_*}\,\Phi^2\!+\! \frac{\lambda}{3}
\Phi^3\!+\!\frac{m^2}{2\,M^3_*}
\Phi\Box\Phi\bigg]\!+\!h.c.\quad
\label{susyint}
\end{eqnarray}
If one keeps all terms in the series expansion above, 
the theory is ghost-free; the effective field 
theory (\ref{susyint}) is valid only below $M_*$.

From this discussion the following question emerges: 
is it possible to reformulate a supersymmetric theory with
higher dimensional operators of class B, in terms of a theory with 
operators of class A only (i.e. with at most two derivatives)?
As we shall see shortly, the answer is in many cases 
affirmative \cite{Antoniadis:2007xc}.
Such a reformulation has interesting advantages. 
The coupling to gravity would become much simpler, and, as a
result supersymmetry breaking is easier to
study. In particular, the coupling of the supersymmetry breaking sector 
to the visible sector can be analysed by the usual standard 
methods. Given the presence of ghost
superfields in the action, one could also ask whether such a theory 
makes sense. The answer is affirmative, as long as one treats the low energy
theory as an {\it effective} theory, valid at energies $E\ll M_*$,
where $M_*$ is essentially the mass of the ghost(s).

\subsection{Higher dimensional operators in the  superpotential}

Let us give an example with operators of class B in the
superpotential. Similar considerations apply to  when
these are present in the K\"ahler potential \cite{Antoniadis:2007xc}.
Consider
\begin{eqnarray}
\!{\cal L}&=&
\int d^4\theta \,\,\Phi^\dagger \Phi
\nonumber\\
&+&\!\!\!\int  d^2\theta\,\bigg[\,
\frac{s}{M_*}\,\Phi\Box\Phi +\frac{m}{2}\,\Phi^2
+\frac{\lambda}{3}\,\Phi^3 \bigg]+h.c.\qquad
\end{eqnarray}
with $s=\pm 1$.  A field redefinition of $\cal L$, 
which treats $\Phi$ and $\Phi'\equiv 
\overline D^2\Phi^\dagger$ as two 
superfields of a Lagrangian with constraints 
(since $\Phi$, $\Phi'$ are not independent), brings $\cal L$ to the form 
\cite{Antoniadis:2007xc}
\begin{eqnarray}\label{unl}
{\cal L}&=&
\int d^4\theta \,\Big[\,{\Phi_1}^{\!\dagger} \Phi_1-
{\Phi_2}^{\!\dagger} \Phi_2\Big]
\nonumber\\
&&
\!\!\!\!\!\!\!\!\!\!\!\!\!\!\!\!\!\!\!\!\!\!\!\!
+\,\,\bigg\{\int d^2\theta \bigg[
\frac{M_*}{16 \,s\sqrt\eta}
\Big( (1-\sqrt\eta)\,\Phi_1-(1+\sqrt\eta) \,\Phi_2\Big)^2
\nonumber\\
&&
\!\!\!\!\!\!\!\!\!\!\!\!\!\!\!\!\!\!\!\!\!\!\!\!
+\,\,\frac{m}{2\sqrt\eta} (\Phi_2-\Phi_1)^2+
\frac{\lambda}{3 \eta^{\frac{3}{4}}} \,(\Phi_2-\Phi_1)^3
\bigg]+h.c.\bigg\}
\end{eqnarray}
where $\eta=1+(17/16)\,m^2/M_*^2$.  
For $m\ll M_*$, $\eta\rightarrow 1$ and 
then (\ref{unl}) simplifies considerably.
The relation between initial fields and new $\Phi_{1,2}$ can
be found in \cite{Antoniadis:2007xc}.
This result is easily extended for a general (derivative-free)
contribution $W(\Phi,\chi)$ to the superpotential, 
present on top of the above class B operator:
\begin{eqnarray}
{\cal L}&=&
\int d^4\theta \Big[\Phi^\dagger \Phi+\chi^\dagger\,\chi\Big]
\nonumber\\
&+&\int  d^2\theta\,\bigg[\,
\frac{s}{M_*}\,\Phi\,\Box\,\Phi +W\big(\Phi;\chi\big) \bigg]+h.c.\qquad
\end{eqnarray}
This 
can  be re-written, if $m\ll M_*$  \cite{Antoniadis:2007xc}:
\begin{eqnarray}
{\cal L}&=&
\int d^4\theta \,\Big[\,{\Phi_1}^{\!\dagger} \Phi_1-
{\Phi_2}^{\!\dagger} \Phi_2+\chi^\dagger \chi \Big]
\nonumber\\
&+&
\int\!\! d^2\theta \bigg[
\frac{M_*}{4\,s}\Phi_2^2 +W\big(\Phi_2-\Phi_1; \chi\big)
\bigg]+h.c.\qquad
\end{eqnarray}
These examples show how to ``unfold'' the original Lagrangian
with operators of class B into a form 
with operators of class A only and  additional superfields ($\Phi_2$).

In these examples the scalar potential takes the form
\begin{equation}
V=\sum_{particles} \vert F_i \vert^2
-\sum_{ghosts}\vert F_j\vert^2
\end{equation}
The first contribution comes from particles and
the second from the ghost degrees of freedom. 
One can then have $V\!>0$, or $V\!<0$, or even
 $V\!=\!0$ with broken supersymmetry. 
The breaking can be done by 
a non-trivial auxiliary field expectation value of
 particle-like $F_i$, ghost-like $F_j$ or of both 
types of fields. Consider for
 example a toy model with explicit soft breaking
\begin{eqnarray}
{\cal L}&=&
\int d^4\theta \Big[
{\Phi_1}^\dagger \Phi_1-
{\Phi_2}^\dagger \Phi_2\Big]\nonumber\\
&+&\!\!
\int d^2\theta \,W\big(\Phi_1-\Phi_2\big)
-m_0^2\,(\phi_1-\phi_2)^2+h.c.\qquad
\end{eqnarray}
where $\phi_{1,2}$ are the scalar fields components of $\Phi_{1,2}$.
The two auxiliary fields have identical eqs of motion, so
 $V(\phi_{1,2})=V_{soft}(\phi_{1,2})$ and $V$ has a minimum
at $\phi_1=\phi_2$. Assuming $W'\not=0$, possible if $W$ contains a
linear term $g(\Phi_1-\Phi_2)$, then $F_1=F_2=g\not=0$, so
 supersymmetry is broken, yet the overall
scalar potential is vanishing.

\subsection{MSSM with higher dimensional operators}

We consider the extension of the MSSM by higher dimensional
operators of class A and B
 and examine their physical effects \cite{Antoniadis1}. 
 Operators of
class A and/or class B are generated 
by integrating out massive superfields which
have interactions with light superfields or which mix with them. 
For example a superpotential with a massive gauge-singlet superfield
$\sigma$,  $W=\lambda\sigma\,H_1\,H_2+M_*\sigma^2$
gives upon integrating out $\sigma$, an effective $W$:
\begin{equation}\label{d51}
W=\frac{\lambda^2}{M_*}\,(H_1\,H_2)^2.
\end{equation}
Another possibility is to have two  massive SU(2) doublet superfields
$H_{3,4}$ which couple to the two MSSM Higgs doublets $H_{1,2}$.
 Ignoring for a moment any gauge interactions, then from
\begin{eqnarray}
{\cal L}\!\!\!\!\!\!&=&
\!\!\!\!\int d^4\theta \bigg[
\sum_{i=1}^4 {H_i}^\dagger H_i\! +\! \big(\nu_1 \,{H_1}^\dagger H_3+
\nu_2 {H_2}^\dagger  H_4 \!+\! h.c.\big)\bigg]
\nonumber\\
&+&
\!\!\!\!\int d^2\theta
\Big[
\mu\,H_1\,H_2+M_*\,H_3\,H_4\Big]+h.c.
\end{eqnarray}
one finds after integrating out $H_{3,4}$
(with $\mu\ll M_*$):
\begin{eqnarray}\label{d52}
{\cal L}\!\!\!&=&\!\!\!
\int d^4\theta \bigg[
{H_1}^\dagger H_1+
{H_2}^\dagger H_2+
\sum_{i=1,2} 
 \frac{\nu_i^2}{M_*^2}\,\,
 \,{H_i}^\dagger\Box  H_i\,\,\bigg]
\nonumber\\
&+&\!\!\!
\int d^2\theta
\bigg[
\mu\,H_1\,H_2+\frac{\nu_1\,\nu_2}{M_*} \,\,
H_1\Box H_2\bigg]+h.c.
\end{eqnarray}
If gauge interactions are present, the last term becomes
\begin{eqnarray}\label{d53}
\frac{\nu_1\nu_2}{4 M_*}\int d^4\theta\,\,\Big[
\,H_2\,e^{-V_1}\, D^2\,e^{V_1}\,H_1+h.c.\,\Big]
\end{eqnarray}
where $V_1\equiv g_2\,\vec V_w\vec \sigma -g_1\,V_Y$.
This Lagrangian contains operators with more than two 
derivatives and can be unfolded
into one with two space-time derivatives
only, as seen above.
This ends our discussion on the origin of these operators
(for details see the Appendix of \cite{Antoniadis1}).

Let us now examine the physical implications
of such operators. We consider an extension of the 
MSSM (hereafter called MSSM$_5$), 
 by all possible $d=5$ operators which respect the R-parity 
symmetry. These  are similar to the operators in (\ref{d51}), 
(\ref{d52}), (\ref{d53}). We ignore the $d=6$ ones since they
are sub-leading. The new Lagrangian
is
\begin{eqnarray}\label{lll}
{\cal L}= {\cal L}_{0}+{\cal L}^{(5)}
\end{eqnarray}
where
\begin{eqnarray}\label{LMSSM}
\!{\cal L}_{0}
&=& 
\!\int d^4\theta \,\Big[\,
{\mathcal Z}_1\,H_1^\dagger \,e^{V_1}\,H_1+
{\cal Z}_2\,H_2^\dagger \,e^{V_2}\,H_2\Big]
 +\cdots 
\nonumber\\
\!\!&+&\!\!\!\!\!\! \bigg\{
\!\int d^2\theta\,\Big[ \,Q\,\lambda_U\,U^c\,H_2
- Q\,\lambda_D\,D^c\,H_1
-  L\,\lambda_E\,E^c\,H_1
\nonumber\\
&&\qquad +\,\,
\mu\,H_1\,H_2
\Big]+h.c.\bigg\}
\end{eqnarray}
The dots in (\ref{LMSSM}) stand for Higgs-independent terms and 
\begin{eqnarray}\label{ll5}
&&\!\!\!\!\!\!\!\!\!\!\!\!\!\!\!\!
{\cal L}^{(5)}=
\frac{1}{M_*}
\int d^4\theta\,\Big[
H_1^\dagger \,e^{V_1} Q\,Y_U\,U^c\,
+H_2^\dagger\,e^{V_2} Q\,Y_D\,D^c\,
\nonumber\\[5pt]
&&\!\!\!\!\!\! \!\!\!\!\!\!
+\quad
{H_2}^\dagger\,e^{V_2} L\,Y_E\,E^c\,+
A\,D^{\alpha}\,\big[\,B\,H_2 \,e^{-V_1}
\big]\,\,D_{\alpha}\,\big[\Gamma\,e^{V_1}\,H_1\,\big]
\nonumber\\[5pt]
 && \!\!\!\!\!\! \!\!\!\!\!\!
+\quad\delta ({\overline\theta}^2)
\big[
 Q\,U^c\,T_Q\,Q\,D^c+ Q\,U^c\,T_L\,L\,E^c+
 \lambda_H
 (H_1 H_2)^2\big]
\nonumber\\[5pt]
&& \!\!\!\!\!\! \!\!\!\!\!\!
+ \quad h.c.\Big]
\label{dim5}
\end{eqnarray}
with $Q, U^c, D^c, L, E^c$ the quark and lepton superfields in a
self-explanatory notation.
Above we introduced the following spurion dependent function-coefficients:
\begin{eqnarray}
A&=& A(S,S^\dagger),\qquad B=B(S,S^\dagger),
\nonumber\\[5pt]
\Gamma&= &\Gamma(S,S^\dagger),\qquad
{\cal Z}_{1,2}={\cal  Z}_{1,2}(S,S^\dagger)
\nonumber\\[5pt]
T_Q&= & T_Q(S),\qquad\quad\,\,\, T_L= T_L(S),
\nonumber\\[5pt]
\lambda_H&= &\lambda_H(S),\,\,
Y_F= Y_F(S,S^\dagger),\quad F=U,D,E\,\,\,\,\,
\end{eqnarray}
where $S=M_s \theta^2$ is the spurion superfield and
$M_s$ the supersymmetry breaking scale.
Any supersymmetry breaking associated with the presence
of the above interactions 
is included using the spurion field technique.

Not all operators in (\ref{dim5}) are independent
\cite{Antoniadis1}.
 To remove this
operator ``redundancy'' we introduce the field re-definitions
\begin{eqnarray}\label{tra}
H_1 &\rightarrow& H_1-\frac{1}{M_*}\,
\overline D^2\,\Big[\Delta_1\,H_2^\dagger\,
e^{V_2}\,(i\,\sigma_2)\Big]^T
+\frac{1}{M_*} \,Q\,\rho_U\,U^c
\nonumber\\[4pt]
H_2 &\rightarrow& H_2
+
 \frac{1}{M_*}\, \overline
D^2\,\Big[\Delta_2\,H_1^\dagger\,e^{V_1}\,(i\sigma_2)\Big]^T
+\frac{1}{M_*}\,Q\,\rho_D\,D^c
\nonumber\\
&+&\frac{1}{M_*}\,L\,\rho_E\,E^c 
\end{eqnarray}
where 
\begin{eqnarray}
\rho_F=\rho_F(S);\,\,\,\,F:U,D,E,\,\,\,
\Delta_i=\Delta_i(S,S^\dagger)\,\,\,\,i=1,2 \label{ft}
\end{eqnarray}
can be chosen arbitrarily.
To avoid the presence of flavour changing 
neutral currents (FCNC),  
the following simple ansatz is made
\begin{eqnarray}\label{ansatz0}
T_Q(S)& = & c_Q(S)\,\,\lambda_U(0)\otimes
\,\lambda_D(0)\nonumber\\[3pt]
T_L(S)& = & c_L(S)\,\,\lambda_U(0)\otimes \,\lambda_E(0)\nonumber\\[3pt]
\rho_F(S)& = & c_F(S)\,\,\lambda_F(0)
\nonumber\\[3pt]
Y_F(S,S^\dagger)&=& y_F(S,S^\dagger)\,\,\lambda_F(0),\,\,
\,\,\, F: U,D,E.
\end{eqnarray}
and, as usual
\begin{eqnarray}\label{lambdas3}
\lambda_F(S)&=&\lambda_F(0)\,(1+A_F \,S),\,\,\,\,\, F: U,D,E.
\end{eqnarray}
With these and a suitable choice for the coefficients of the
spurion  entering in $\Delta_{1,2}$
one can set $T_{Q}=T_L=0$ also $A=B=\Gamma=0$
and $Y_F\rightarrow y_F(S^\dagger)\,\lambda_F(0)$, $F=U,D,E$.
Then ${\cal L}^{(5)}$ becomes
\begin{eqnarray}\label{lastL}
{\cal L}^{(5)}\!\!\!\!\!\!\!\!\!
&=&\!\! \frac{1}{M_*}\!
\,\int\! d^4\theta\,\Big[
H_1^\dagger\,e^{V_1}\,Q\,Y_U'(S^{\dagger})\,U^c
\nonumber\\[3pt]
&+&
H_2^\dagger\,e^{V_2} Q\,Y_D'(S^{\dagger})\,D^c
+
H_2^\dagger\,e^{V_2}  L\,Y_E'(S^{\dagger})\,E^c+h.c.\Big]
\nonumber\\[3pt]
&+&\frac{1}{M_*}\int
d^2\theta\,\,\lambda_H'(S)\,(H_1\,H_2)^2+h.c.
\label{final2}
\end{eqnarray} 
with $V_2\equiv g_2\,\vec V_w\vec \sigma +g_1\,V_Y$.
The new Yukawa couplings $Y_F^{'}(S^\dagger)$, $F:U,D,E$
have now a dependence on $S^\dagger$ only:
\begin{eqnarray}\label{ynew}
Y_F^{'}(S^\dagger)=\lambda_F(0)\,(x_0^F+x_2^F\,S^\dagger)
\end{eqnarray}
Following (\ref{tra}), the couplings of ${\cal L}_0$ (\ref{LMSSM}) and
${\cal Z}_{1,2}$ have acquired, at the {\it classical level},
threshold corrections which depend on the scale $M_*$ 
 \cite{Antoniadis1}.
The new form of ${\cal L}^{(5)}$ in (\ref{lastL}) 
gives the minimal irreducible set of
R-parity conserving dimension-five  operators that can be present
beyond MSSM.

\subsubsection{Physical consequences: corrections to the Higgs mass}

Let us address some of the physical consequences of the 
higher dimensional operators in ${\cal L}^{(5)}$ of (\ref{lastL}).
For related studies see \cite{Dine,Brignole,Pospelov,Blum}.
The Higgs scalar potential $V_H$ obtained from (\ref{lastL}) is:
\begin{eqnarray}\label{scalarV}
V_H&=&\tilde m_1^2\,\vert h_1\,\vert^2
+\tilde m_2^2\,\vert h_2\,\vert^2
+\big( \,B\, \mu \,h_1 \,h_2+h.c.\big)
\nonumber\\
&+&\frac{g^2}{8}\,\big(\vert \,h_1\,\vert^2-\vert\,
h_2\,\vert^2\big)^2+
\frac{1}{2}\,\big(\,\eta_3\,(h_1\,h_2)^2+h.c.\big)
\nonumber\\[1pt]
&+&
 \big(\vert\,h_1\,\vert^2
-\vert\,h_2\,\vert^2\big)\,\big(\eta_1 \,h_1\,h_2+h.c.\big)
\nonumber\\[3pt]
&+&
\big(\vert\,h_1\,\vert^2+\vert\,h_2\,\vert^2\big)\,
\big(\eta_2 \,h_1\,h_2+h.c.\big)
\end{eqnarray}
where $g^2=g_2^2+g_1^2$.
Here $\eta_1\propto g^2\,M_s/M_*$, $\eta_2\propto 2\mu/M_*$, 
$\eta_3\propto M_s/M_*$. $\eta_1$ is due to the  derivative 
term in (\ref{dim5}). Although its contribution to $V_H$ can be
removed by redefinitions (\ref{tra}), up to a finite
renormalisation of the soft masses, one can however  keep it, 
in order to see the effects of  such renormalisation. 

In  the limit of large $\tan\beta=v_2/v_1$ with the pseudoscalar mass parameter $m_A$ 
fixed at a value $m_A>m_Z$, one finds:
\begin{eqnarray}
\label{rr1}
m_h^2=m_Z^2+\frac{4 m_A^2 \,{  v}^2}{m_A^2-m_Z^2}\,(\eta_2-\eta_1)
\,\cot  \beta+{\cal O}(\cot^2\beta)
\end{eqnarray}
where $v^2={v_1^2+v_2^2}$.
This would suggest that an increase of the mass of the lightest 
Higgs above $m_Z$ would be possible, thus lifting the tree level bound
we have in the MSSM. However, this expansion  valid at large
$\tan\beta$ only, breaks perturbative expansion in $1/M_*$ since
then  dimension-six operators and higher are relevant. 
Note that $\eta_1$ plays no role in the
relation among physical masses since 
\begin{eqnarray}
m_H^2+m_h^2=m_A^2+m_Z^2
+2\eta_2\,v^2\,\sin2\beta+\eta_3\,v^2
\end{eqnarray}
A numerical analysis shows that the lightest Higgs mass can be
increased mildly relative to $m_Z$ by up to $\epsilon_r= 16 \%$ 
($m_h\leq 105$ GeV) for $m_A$ close to $m_Z$; however if $m_A$
 increases above $m_Z$, $\epsilon_r$  is very small.
In conclusion, quantum corrections are still needed for a value
of $m_h$ above the bound of $114$ GeV, like in the MSSM; however,
in the MSSM$_5$ the amount of stop mixing needed to achieve this can be
relaxed relative to the MSSM case. In conclusion the MSSM Higgs sector
is rather stable under the addition of higher dimensional operators,
in our approximation of including only $d=5$ operators.

\subsubsection{Physical consequences: new couplings
 from ${\cal L}^{(5)}$.}

Another consequence of the presence of the irreducible set of
dimension-five operators in (\ref{lastL}) is 
the generation of new couplings, beyond those present in the  MSSM at
 the tree level.
One new coupling  is a 
 ``wrong''-Higgs Yukawa coupling,  which exchanges usual holomorphic
dependence on one Higgs by the dependence on the hermitian conjugate
of the other ($H_1\leftrightarrow H_2^\dagger$) \cite{Haber,
  M0}. Such couplings can also arise in the MSSM at one loop, upon
integrating out massive squarks, where they are suppressed by
 $M_s^2/M^2_*\times\,\,(loop-factor)$. 
Here they are suppressed by $M_s/M_*$ only, as seen below:
\begin{eqnarray}\label{cset3}
&& \frac{M_s}{M_*}\,x_2^U\,(\lambda^U_0)_{ij}\,\,
(h_1^\dagger\,q_{L\,i})\,\,u_{R\,j}^c+h.c.\nonumber\\
&&  \frac{M_s}{M_*}\,x_2^D\,(\lambda^D_0 )_{ij}\,\,
(h_2^\dagger\,q_{L\,i})\,\,d_{R\,j}^c+h.c.\nonumber\\
&& \frac{M_s}{M_*}\,x_2^E\,(\lambda^E_0 )_{ij}\,\,
 (h_2^\dagger\,l_{L\,i})\,\,e_{R\,j}^c+h.c.,
\end{eqnarray}
with the notation:
$\lambda^F_0\equiv \lambda_F(0),\,\,\, F:U,D,E$,
and where $x_2^F$ can be read from (\ref{ynew}).
These couplings can bring a $\tan\beta$ enhancement of a prediction
for a physical observable, such as the bottom quark mass, relative to
bottom quark Yukawa coupling:
\begin{eqnarray}
m_b=\frac{ v \cos\beta}{\sqrt
  2}\,\Big( \lambda_b +{\delta}{\lambda_b}+
{\Delta}{\lambda_b}\tan\beta\Big)
\end{eqnarray}
Here $\lambda_b$ is the ordinary bottom quark Yukawa coupling, 
$\delta\lambda_b$  is its one-loop correction in the MSSM and
$\Delta\lambda_b$ is a ``wrong''-Higgs coupling' contribution,
obtained from integrating our massive squarks at one loop in the MSSM,
which in our case receives an additional correction from (\ref{cset3}).
This last correction can be larger than its one-loop-generated MSSM
counterpart \cite{Haber,C1,C2,C3}. 
This can bring a $\tan\beta$ enhancement of
the Higgs decay rate into bottom quarks pairs.

Note that in the MSSM$_5$  defined by eq.(\ref{lastL}),
couplings proportional to $M_s$ involving ``wrong''-Higgs A-terms
are not present, given our FCNC ansatz (\ref{ansatz0}) leading to
(\ref{ynew}).
 If this ansatz is not imposed on the third generation,
then  one could have such terms:
\begin{eqnarray}
\frac{M_s^2}{M_*}\Big[
y_{u,3}\,
 h_1^\dagger\,\tilde q_{L,3}\tilde u_{R,3}^*
+
y_{d,3}\, 
h_2^\dagger\,\tilde q_{L,3}\tilde d_{R,3}^* 
+
 y_{e,3}\,
h_2^\dagger\,\tilde l_{L,3}\tilde e_{R,3}^* \Big]
\!\!\!\!\!\!\!\!\!
\nonumber
\end{eqnarray}
where $y_{f,3}$, $f=u,d,e$ are the coefficients of
the component $S\,S^\dagger$ of $Y'(S,S^\dagger)$ of the 
third generation.

There are also new, important
 supersymmetric couplings that are generated, which
affect the amplitude of processes like
quark + quark $\rightarrow$ squark + squark,
 or involving (s)leptons as well.
These are
\begin{eqnarray}\label{qqsqsq} 
&&\!\!\!\!\!\!\!\!\!\!\!\!\!\!\!\!\!\!
\frac{x_0^U}{M_*}\,(\lambda_0^D)_{ij}\,(\lambda_0^U)_{kl}
\,\,\tilde q_{L\,i}\,\tilde d_{R\,j}^*\,\,
q_{L\,k}\,u_{R\,l}^c+ h.c.
\nonumber\\
 &&\!\!\!\!\!\!\!\!\!\!\!\!\!\!\!\!\!\!
 \frac{x_0^D}{M_*}\,(\lambda_0^U)_{ij}\,(\lambda_0^D)_{kl}
 \,\,\tilde q_{L\,i}\,\tilde u_{R\,j}^*\,\, q_{L\,k}\,d_{R\,l}^c+h.c.
\\
&&\!\!\!\!\!\!\!\!\!\!\!\!\!\!\!\!\!\!
\frac{x_0^U}{M_*}\,(\lambda_0^E)_{ij}\,(\lambda_0^U)_{kl}
\,\,\tilde l_{L\,i}\,\tilde e_{R\,j}^*\,\, q_{L\,k}\,u_{R\,l}^c
+(L\!\leftrightarrow\! Q, E\!\leftrightarrow \!U)\!+\!h.c.\nonumber
\end{eqnarray}
These couplings
can be important particularly for the third generation.
The largest effect would be for squarks pair production
from a pair of quarks; the corresponding amplitude 
can be comparable to the MSSM tree level
 contribution \cite{Dawson,Harrison}. Consider for example $q {\bar q}
\rightarrow {\tilde q} {\tilde q^*}$, generated by a tree-level
gluon exchange. The MSSM amplitude behaves as the first term in
$A^{total}_{q {\bar q} \rightarrow g \rightarrow {\tilde q} {\tilde q^*}}$:
\begin{equation}
A^{total}
_{q {\bar q} \rightarrow g \rightarrow {\tilde q} {\tilde q^*}} \sim
{\frac{g_3^2}{\sqrt{s}}} \ 
+\frac{\lambda_0^U \lambda_0^D}{M_*} 
\end{equation}
where $\sqrt s$ is the center of mass energy.
The second term
is generated by the additional  couplings in (\ref{qqsqsq}).
While the MSSM contribution decreases with $s$, the second term 
is constant with potentially significant effects.

\vspace{-0.1cm}
\subsection{Conclusions}

Effective field theories provide an appropriate framework for the
study of new physics beyond the SM and the MSSM.
In these theories our ignorance of high energy physics 
is parametrised in terms of higher dimensional operators, which
are organised in terms of inverse  powers of the scale 
of new physics $M_*$. To further restrict
the exact form of the effective action,
other organising criteria are used, such as
symmetry principles inspired by the low energy phenomenology.
Using these criteria, one is then able to make
 testable predictions for the low energy observables.
This is important since often the exact details 
of the high-scale, fundamental theory are not known in detail
(moduli problem, vacua degeneracy, etc). It is difficult 
to make testable predictions in this,  and the use of 
effective theories can provide a successful approach.

There are two classes of higher dimensional operators,
with up to two space-time derivatives (class A) and with more than two such
derivatives (class B). While the former class
is more studied, class B is also a common presence. Class B operators are 
generated in 4D renormalisable theories, by
integrating  massive fields, with the result of generating an
infinite series of derivatives.  Truncating this series can generate
ghosts fields, which signals that the theory is only valid below the
scale of these states. Using general field redefinitions one can
reformulate a theory with both classes of operators in terms of a
second-order one with class A operators only. This can have 
applications when coupling  such theories to gravity.

We considered the study  of the R-parity conserving, dimension-five operators
and their generalisation to the supersymmetry breaking case,
that extend the MSSM Lagrangian. Not all these operators are
independent. Using general, spurion dependent
field redefinitions, we removed the redundancy and
identified the minimal irreducible set
of dimension-five operators that can exist beyond the MSSM.
The phenomenological implications of this MSSM extension were studied.
It turns out that the MSSM Higgs sector is rather stable, in the
approximation used, under the presence of these operators, although
a mild increase of the lightest neutral Higgs scalar may be present,
up to $\approx 105$ GeV, before 
taking into account quantum corrections.
Additional couplings are also generated, and can dominate
their counterparts generated in the MSSM alone 
at the loop-level. For example squark
production and Higgs decays into b-quarks are
enhanced by the presence of dimension-five operators.
 The method to
identify the irreducible set of higher dimensional operators
is general and can be applied to other models, too.

\vspace{-0.1cm}
\begin{theacknowledgments}
This work was partially supported by ANR (CNRS-USAR) contract
05-BLAN-007901,   INTAS grant 03-51-6346,
EC contracts MRTN-CT-2004-005104, 
MRTN-CT-2004-503369 and
MRTN-CT-2006-035863,  CNRS PICS
\#~2530,  3059, 3747, 4172,  and  European Union Excellence Grant
MEXT-CT-2003-509661.
\end{theacknowledgments}


\vspace{-0.1cm}
\begin{thebibliography}{9}

\bibitem{Hawking}
S.W. Hawking, T. Hertog, \emph{Phys. Rev. D} {\bf 65} 103515 (2002)


\bibitem{Antoniadis:2007xc}
  I.~Antoniadis, E.~Dudas and D.~M.~Ghilencea,
  \emph{JHEP} {\bf 0803} 045 (2008)
  [arXiv:0708.0383 [hep-th]].

\bibitem{NH1}
S.~Groot Nibbelink and M.~Hillenbach,
 \emph{Phys.\ Lett.\ B} {\bf 616} 125-134 (2005)
  [arXiv:hep-th/0503153];

\bibitem{NH2}
 S.~Groot Nibbelink and M.~Hillenbach,
\emph{Nucl. Phys. B} {\bf  748} 60-97 (2006). [arXiv:hep-th/0602155].


\bibitem{Gh0}
  D.~M.~Ghilencea, H.~M.~Lee and K.~Schmidt-Hoberg,
  JHEP {\bf 0608}  009 (2006)
  [arXiv:hep-ph/0604215].

\bibitem{Gh1}
  D.~M.~Ghilencea,
 \emph{JHEP} {\bf 0503} 009 (2005) 

\bibitem{Gh2}
  D.~M.~Ghilencea and H.~M.~Lee,
  JHEP {\bf 0509}  024 (2005)

\bibitem{Antoniadis1}
  I.~Antoniadis, E.~Dudas, D.~M.~Ghilencea, P.~Tziveloglou,
  [arXiv:0806.3778 [hep-th]] (submitted to \emph{Nucl.\ Phys.\  B}).


\bibitem{Dine}
  M.~Dine, N.~Seiberg and S.~Thomas,
  \emph{Phys.\ Rev.\  D} {\bf 76} 095004 (2007)
  [arXiv:0707.0005 [hep-ph]].

\bibitem{Brignole}
  A.~Brignole, J.~A.~Casas, J.~R.~Espinosa and I.~Navarro,
\emph{Nucl.\ Phys.\  B} {\bf 666} 105-143 (2003)
  [arXiv:hep-ph/0301121].


\bibitem{Pospelov}
  M.~Pospelov, A.~Ritz and Y.~Santoso,
  \emph{Phys.\ Rev.\  D} {\bf 74} 075006 (2006)
  [arXiv:hep-ph/0608269].

\bibitem{Blum}
  K.~Blum and Y.~Nir,
 \emph{Phys.\ Rev.\  D} {\bf 78} 035005 (2008)
  [arXiv:0805.0097 [hep-ph]].

\bibitem{Haber}
  H.~E.~Haber, J.~D.~Mason,
\emph{  Phys.\ Rev.\  D} {\bf 77} 115011 (2008)
[arXiv:0711.2890 [hep-ph]].

\bibitem{M0}
  S.~P.~Martin,
\emph{  Phys.\ Rev.\  D} {\bf 61} 035004 (2000)

\bibitem{C1}
  M.~S.~Carena, H.~E.~Haber, H.~E.~Logan and S.~Mrenna,
  \emph{Phys.\ Rev.\  D} {\bf 65} 055005 (2002).

\bibitem{C2}
D.~M.~Pierce, J.~A.~Bagger, K.~T.~Matchev and R.~J.~Zhang,
\emph{Nucl.\ Phys.\  B} {\bf 491} 3-67 (1997)
[arXiv:hep-ph/9606211].

 \bibitem{C3}
 L.~J.~Hall, R.~Rattazzi and U.~Sarid,
\emph{Phys.\ Rev.\  D} {\bf 50} 7048-7065 (1994)
[arXiv:hep-ph/9306309].

\bibitem{Dawson}
  S.~Dawson, E.~Eichten and C.~Quigg,
  \emph{Phys.\ Rev.\  D} {\bf 31} 1581 (1985).

\bibitem{Harrison}
  P.~R.~Harrison and C.~H.~Llewellyn Smith,
\emph{  Nucl.\ Phys.\  B} {\bf 213} 223 (1983).
  [Erratum-ibid.\  B {\bf 223} 542 (1983)].

\end{thebibliography}
\end{document}